\documentclass[showpacs,twocolumn,pre]{revtex4}
\usepackage{graphicx}

\newcommand{\boldmu}{\mbox{\boldmath $\mu$}}
\newcommand{\bolddelta}{\mbox{\boldmath $\Delta$}}
\newcommand{\erf}{\mathrm{erf}}
\newcommand{\kBT}{k_\mathrm{B}T}
\newcommand{\Eb}{E_\mathrm{b}}
\newcommand{\xb}{x_\mathrm{b}}

\newcommand{\vmin}{v_\mathrm{min}}
\newcommand{\vmax}{v_\mathrm{max}}

\begin{document}

\title{Single-molecule force spectroscopy: 
Practical limitations beyond Bell's model}
\author{Sebastian Getfert}
\author{Mykhaylo Evstigneev}
\author{Peter Reimann}
\affiliation{Universit\"at Bielefeld, Fakult\"at f\"ur Physik, 33615
Bielefeld, Germany}
 
\begin{abstract}
Single-molecule force spectroscopy experiments, as well as
a number of other physical systems, are governed by 
thermally activated transitions out of a metastable state 
under the action of a steadily increasing external force.
The main observable in such experiments is the
distribution of the forces, at which the escape events occur. The
challenge in interpreting the experimental data is to relate them
to the microscopic system properties.
We work out a maximum likelihood approach and show that 
it is the optimal method to tackle this problem. 
When fitting actual experimental data it is unavoidable 
to assume some functional form for the force-dependent 
escape rate. 
We consider a quite general and common such functional form and
demonstrate by means of data from a realistic computer 
experiment that the maximum number of fit parameters 
that can be determined reliably is three. 
They are related to the force-free escape rate
and the position and height of the activation
barrier. 
Furthermore, the results for the first two of these fit
parameters show little dependence on the assumption about the manner
in which the barrier decreases with the applied force, while
the last one, the barrier height in the absence of force, 
depends strongly on this assumption.
\end{abstract}
                                                 

\maketitle

\section{Introduction}
\label{introduction}
A quite remarkable experimental achievement of the last decade
is the direct observation of chemical dissociation at the 
single-molecule level by applying time-dependent
external forces on the pico-Newton scale. This technique is
called dynamic force spectroscopy or single-molecule 
force spectroscopy and reviewed e.g. in
\cite{Mer01, Eva01}.
It allows to extract kinetic constants and
energy landscape parameters of various interactions like
antibody-antigen recognition \cite{Hin96} or protein-DNA 
interactions \cite{Bar07}.
Also the dynamics of various other experimental systems are 
governed by thermally activated transitions out of a metastable
state over a potential barrier, whose height decreases in time 
due to a steadily increasing external force.
Examples include the polarization reversal in nanomagnets \cite{Wer97}, 
friction at the atomic scale \cite{Rie03, Sch05, Evs06}, 
and phase jumps in Josephson junctions
\cite{Kur72, Ful74}. 
In all these cases, the knowledge of the force-dependent rate 
out of the metastable state can be exploited to characterize 
the system studied. 
The main theme of the present work is how to perform such a 
characterization in the most optimal way and to point out the 
limitations even of such an optimized  
procedure under practical conditions.

A typical experimental setup is schematically sketched in Fig. 1: the
single chemical bond of interest, e.g. in a ligand-receptor complex,
is connected via two linker molecules with the tip of an AFM (atomic
force microscope) cantilever (or some other micromechanical tool) and
a piezoelectric element.  The latter is employed for ``pulling down''
the attached linker molecule at some constant velocity, leading to an
elastic reaction force of the cantilever, determined from the
deflection of a laser beam.  The main quantity of interest is the
magnitude of the force at the moment when the bond breaks.

\begin{figure}
  \centering
  \includegraphics[width=1\linewidth]{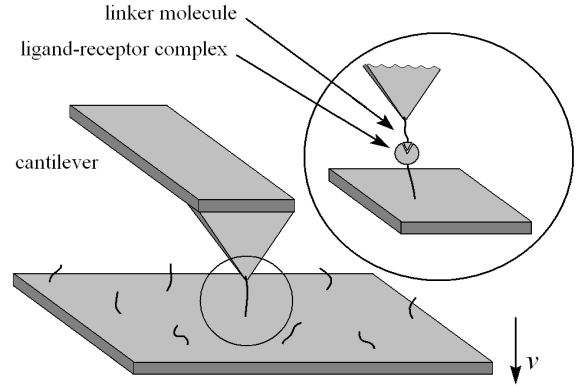}
  \caption{    (Color online)
    Schematic illustration of a single-molecule force spectroscopy
    experiment.  Receptor and ligand are connected via suitable
    linkers to the surface and the AFM tip, respectively. The distance
    of the tip form the surface can be controlled with a piezoelectric
    element (not shown). When pulled down at constant speed an
    (approximately linearly) increasing force acts on the bond which
    can be measured through the deflection of the cantilever.}
  \label{fig1}
\end{figure}

The theoretical interpretation of the observed rupture forces is a
non-trivial task for the following reasons.  Upon repeating the same
experiment with the same pulling velocity, the rupture forces are
found to be distributed over a wide range, contrary to what one would
naively expect for a purely mechanical breaking of a compound object
at some fixed, ``critical'' strain force.  A further theoretical
challenge represents the experimental finding that for different
pulling velocities different such distributions are obtained.  Hence,
neither a single rupture event nor the average rupture force at any
fixed pulling velocity can serve as a meaningful characteristics of a
given chemical bond strength.  Major steps in solving the puzzle are
due to Bell \cite{Bel78} and to Evans and Ritchie \cite{Eva97},
recognizing that a forced bond rupture event is a thermally activated
decay of a metastable state that can be described within the general
framework of Kramers reaction rate theory \cite{Han90}.  Subsequently,
their basic theoretical approach has been extended and refined in
several important directions, see e.g. Refs. \cite{Mer01, Eva01,
Rie98, Mer99, Str00, Hey00, Sei00, Isr01, Bar02, Ngu03, Dud03, Hum03,
Mal06, Dud06, Han06, Hus08}.

Following Evans and Ritchie \cite{Eva97, Mer01},
a single-molecule dissociation process
is viewed as thermally activated escape 
of a reaction coordinate $x$ over a potential 
barrier, see Fig. 2.
Given the one-dimensional potential landscape along the reaction
coordinate, the dissociation rate $k(f)$ for an instantaneous force $f$ 
(projected onto the direction of the reaction coordinate) can
be written according to Kramers reaction-rate 
theory \cite{Han90}
in the form
\begin{equation}
k(f)=\omega(f)\, \exp(- E_b(f)/k_BT), \
\label{1.20}
\end{equation}
where the pre-exponential factor $\omega (f)$ has the intuitive meaning 
of an ``attempt frequency'' and the exponentially leading 
Boltzmann-Arrhenius factor
contains the relevant potential barrier $ E_b(f)$ against escape,
Boltzmann's constant $k_B$, and the temperature $T$.

\begin{figure}
  \centering
  \includegraphics[width=0.75\linewidth]{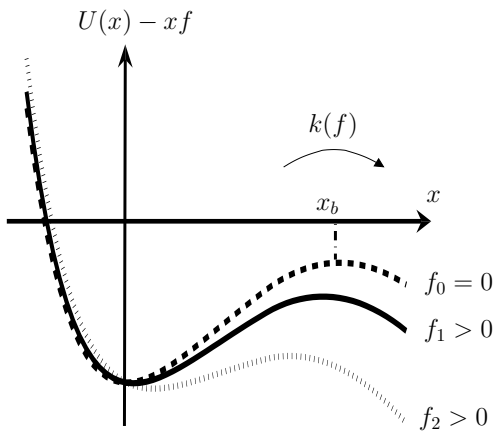}
  \caption{Schematic illustration of the
    total relevant potential energy landscape $U(x)-xf$ 
    of a receptor-ligand-bond as
    function of the reaction coordinate $x$ without 
    and with externally applied bias forces $f_1<f_2$. 
    For low forces
    the main effect is that the energy barrier $\Eb(f)$
    is lowered by an amount $\Delta \Eb\simeq \xb  f$,
    where $\xb $ is the distance between potential well 
    and barrier at zero force.
    For larger forces this distance decreases.   }
  \label{fig2}
\end{figure}

In dynamic force spectroscopy experiments, the rate at which the force
increases is much slower than all other relevant molecular relaxation
processes.  Due to this separation of time scales \cite{Han90}, the
reaction kinetics can be very accurately approximated by the following
first-order differential equation
\begin{eqnarray}
  \frac{dn(t)}{dt} = -k\left(f(t)\right)n(t)\ ,
  \label{1.10}
\end{eqnarray}
where $n(t)$ denotes the survival probability of the bond up to time $t$ and
$k(f(t))$ is the accompanying dissociation rate (\ref{1.20})
at an instantaneous external force $f(t)$.

Ideally, one would wish to use the experimentally established
force-dependent escape rate $k(f)$ to reconstruct the potential
landscape, i.e. potential energy vs. the reaction coordinate,
cf. Fig.~\ref{fig2}.  However, in view of the fact that the escape
rate depends on the energy \emph{difference} at two force-dependent
extrema, this problem does not have a unique solution
\cite{Izr97}. Therefore, one may start with some model energy
landscape, and try to deduce its global features, such as barrier
height in the absence of the force and dissociation length (the
distance between potential well and barrier without bias force,
cf. Fig.~\ref{fig2}). This introduces some specific functional form
for the escape rate (\ref{1.20}) involving several parameters, which
are then determined by fitting the experimental data. In the present
work, we describe the application of the maximum likelihood approach
\cite{Get07, Dud07} as a tool to deduce the model parameters. We show
that this method is superior to any other approach that may be used
for this purpose.

After introducing the method, we discuss its application to determine
the parameters of a commonly used Ansatz for the rate due to Bell,
henceforth called Bell's model \cite{Bel78,Eva97}.  This approximation
assumes that the force is sufficiently small, so that the
force-dependent barrier height decreases linearly with the force, the
proportionality constant $\xb$ being the dissociation length (see
Fig.~\ref{fig2}),
\begin{equation}
\Eb(f) = \Eb(0) - \xb  f 
\ .
\label{1.30}
\end{equation} 
The two fit parameters used in this approach are the escape rate at
zero force, $k_0 = \omega(0)e^{-\Eb(0)/\kBT}$, and the dissociation 
length, $\xb$.  
Within Bell's model, it is impossible to uncouple
the intrinsic time and energy scales of the system, because
multiplication of $\omega(0)$ by an arbitrary constant and addition of
$\kBT$ times the logarithm of that constant to $\Eb(0)$ leaves the
force-free rate value, $k_0$, and hence the statistics of escape
events, the same.

One of the central problems of the present work is gaining more
information about the system than Bell's model
allows. Obviously, in order to make it possible, one needs to use a
functional form of the escape rate $k(f)$ involving more fit
parameters than in Bell's Ansatz. Indeed, it has been suggested
\cite{Dud07, Dud06, Sch05, Evs06, Rie03, Han06} that one can actually
estimate the value of the force-free activation barrier $\Eb(0)$ if
one makes a more realistic assumption about how the barrier decreases
from this value with increasing force; the linear decrease 
(\ref{1.30}) is
replaced with a non-linear potential. Then, the force-dependent rate
involves more than two fit parameters, the force-free barrier height
$\Eb(0)$ being one of them.

It is intuitively clear that the more rate parameters one has, the
better one can characterize the system of interest. On the other hand,
if the number of fit parameters is too large, then not all of them
may be determined sufficiently accurately from the experimental
data. Therefore, a relevant question is: what is the maximal number of
rate parameters, which one can establish reliably? 
For a quite common and general parametrization of the potential 
we show that the highest number of model parameters one can determine from
fitting the experimental data is three, while inclusion of additional
parameters into the theory does not improve the quality of the
fit. These three parameters are related to the force-free escape rate,
the height of the activation barrier, and the dissociation length. 
Furthermore,
we show that the resulting fit values of these parameters strongly
depend on the assumption concerning the manner in which the activation
barrier decreases with the applied force. In other words, one cannot
determine them uniquely without having this information.

\section{Theoretical modeling of bond rupture}
\label{s2}
For a quantitative analysis of dynamic
force spectroscopy experiments one usually assumes that the force
$f(t)$ increases linearly,
\begin{eqnarray}
    f(t) = \kappa v t\ ,
    \label{2.10}
\end{eqnarray}
where $v$ is the (constant) pulling velocity, and $\kappa$ the
relevant total elasticity of cantilever, linker molecules, receptor
and ligand, cf. Fig. 1.  
Using the fact that the extension $s$ of the complex at
time $t$ is $s=vt$ we see that (\ref{2.10}) is equivalent to the
assumption of a linear force-extension characteristics.

Combining Eqs.~(\ref{1.10}) and (\ref{2.10}), we arrive at the central
experimental quantity, namely the probability density of escape events
at the force value $f$ for a given velocity $v$ and a given set of
parameters $\boldmu$ characterizing the escape rate $k(f)$:
\begin{eqnarray}
& & p_1(f | \boldmu, v) = -\frac{dn(f)}{df} = \frac{k(f)}{\kappa v}
e^{-g(f)/\kappa v}\ ,
\label{2.20}
\\ 
& & g(f) := \int_0^f df' k(f')\ .
\label{2.21}
\end{eqnarray}
For later convenience, the rupture force distribution (\ref{2.20})
is written as a
conditional probability, conditioned on the values of the model
parameters $\boldmu$ and the pulling velocity $v$.  
While, in practice, the latter is directly accessible from the
measurement, the model parameters $\boldmu$ have to be inferred from the
distribution of rupture forces.

It is possible to evaluate the integral (\ref{2.21}) numerically for
any functional form of the rate $k(f)$. However, the exponentially
increasing character of this function, see Eq.~(\ref{1.20}), allows one
to derive a very efficient analytical approximation for the
integral.
Namely, in view of Eq.~(\ref{1.20}), the main contribution to
the integral $g(f)$ comes from the $f'$-region just below $f$. This
allows one to expand $\ln k(f')$ in the vicinity of the upper limit of
integration $f$ to the first order  \cite{Hey00}:
\begin{eqnarray}
& &k(f') = k(f)\,e^{\lambda_1(f)(f'-f)}\ , 
\label{2.30}
\\ 
& &\lambda_n(f) := (-1)^{n-1}\frac{d^n\ln k(f)}{df^n}\ .
\label{2.31}
\end{eqnarray}
With this approximation, the integral in (\ref{2.21})
is given by
\begin{equation}
g(f) \approx g_1(f) = \frac{k(f)}{\lambda_1(f)}\left(1 -
  e^{-\lambda_1(f)f}\right)\ .
\label{2.40}
\end{equation}
This simple approximate formula may be sufficient for most practical
purposes. However, if the deviations of $\ln k(f')$ from linearity are
important, one can use the second-order approximation for this
function near $f' = f$:
\begin{equation}
k(f') = k(f)\,e^{\lambda_1(f)(f'-f) - \lambda_2(f)(f'-f)^2/2}\ ,
\label{2.50}
\end{equation}
allowing one to evaluate the rate integral in (\ref{2.21})
as
\begin{eqnarray}
\label{2.60}
& & g(f) \approx g_2(f) =
k(f)e^{\lambda_1(f)^2/[2\lambda_2(f)]}\sqrt{\frac{\pi}{2\lambda_2(f)}}\\
& & \Bigg[\erf\left(\sqrt{\frac{\lambda_2(f)}{2}}
\left(f + \frac{\lambda_1(f)}{\lambda_2(f)}\right)\right)
-\erf\left(\frac{\lambda_1(f)}{\sqrt{2\lambda_2(f)}}\right)\Bigg]\
, 
\nonumber
\end{eqnarray}
where $\erf (x):=2\pi^{-1/2}\int_0^x dy\,  e^{-y^2}$ is the error function.
Additional numerical analysis has shown that the inaccuracy of this
expression is smaller than 1\% for all reasonable functional choices
$k(f)$ which we have checked.
%
%

\section{Parameter Estimation: Properties of the maximum likelihood
estimator}
\label{estimation}
Let us assume that a specific model and thus the particular form of the
probability density $p_1$ in (\ref{2.20}) can be considered as given.  
Then the remaining task is to estimate the
model parameters $\boldmu$ from a given set of $N$ rupture forces
$\mathbf{f} = \{f_i\}_{i=1}^N$ and pulling velocities $\mathbf{v} =
\{v_i\}_{i=1}^N$.  There exist different ``recipes'' for doing this,
called estimators.  Each of them can be formally represented by some
function $\tilde{\boldmu}(\mathbf{f})$, indicating that the estimate,
being a function of the random variables $\mathbf{f}$, 
is a random variable itself.
Now, the main question is: what is the optimal estimate of the model
parameters that can be extracted from the given set of $N$ rupture forces?
Stated differently: which recipe yields estimates
$\tilde{\boldmu}(\mathbf{f})$ of the model parameters which are {\it
on average} over many data sets $\mathbf{f}$ closest to the {\it
``true'' model parameters}?
    
In this section we discuss some properties of estimators and show that
under realistic experimental conditions, given in single-molecule
pulling experiments, no estimator $\tilde{\boldmu}(\mathbf{f})$
outperforms the maximum likelihood estimator.  The reader who is not
interested in the mathematical details may skip
the subsequent subsections A-C and immediately proceed to
subsection D, where we summarize the main steps which are
necessary for a practical application.

\subsection{Maximum Likelihood Estimator}
\label{maxlike}
Our starting point is the probability $p$ to observe a given set of
$N$ rupture forces $\mathbf{f} = \{f_i\}_{i=1}^N$ measured at pulling
velocities $\mathbf{v} = \{v_i\}_{i=1}^N$.  Since the $f_i$ are
statistically independent, this probability reads:
\begin{equation}
p({\bf f} | \boldmu, {\bf v}) = \prod_{i=1}^N p_1(f_i|\boldmu,v_i) \ .
\label{3.10}
\end{equation}
The proceeding consists in simply maximizing (\ref{3.10}) with respect
to $\boldmu$ \cite{Get07, Dud07}; usually, this has to be done
numerically.  For any given $\bf{f}$ and $\bf{v}$ the corresponding
set of parameters $\boldmu^\ast = \boldmu^\ast (\bf{f},\bf{v})$ is
called the {\it maximum likelihood estimate}.

Intuitively, the properties of this estimator are most 
easily understood within the framework of Bayesian inference 
\cite{Ago03,Dos03}.  
The quantity in
(\ref{3.10}) is called {\it likelihood} and plays a central role in the
Bayesian approach. Extending the notion of ``probability'' in the
sense of ``degree of belief'' to the model parameters $\boldmu$, the
joint probability $p({\bf f},\boldmu,{\bf v})$ can be written in terms
of conditional probabilities $p(...|...)$ either in the form
$p(\boldmu|{\bf f},{\bf v})\, p({\bf f},{\bf v})$ or in the form
$p({\bf f}|\boldmu,{\bf v})\, p(\boldmu,{\bf v})$, yielding Bayes'
theorem:
\begin{eqnarray}
  p(\boldmu | {\bf f},{\bf v}) =p({\bf f} | \boldmu,{\bf v})\,
  p(\boldmu, {\bf v})\, [p({\bf f}, {\bf v})]^{-1} \ .
  \label{3.20}
\end{eqnarray}
The left hand side represents the ``likeliness'' of $\boldmu$, given
the data ${\bf f}$, ${\bf v}$, and hence is clearly of central
interest for our purposes.  Considering also the right hand side as a
function of $\boldmu$, it is equal to the likelihood from (\ref{3.10})
times the so called {\it prior probability} $p(\boldmu, {\bf v})$,
encapsulating all our knowledge about $\boldmu$ before the
measurement, times a $\boldmu$-independent factor.  Thus, determining
$p(\boldmu | {\bf f},{\bf v})$ by means of the right hand side of
(\ref{3.20}) provides the central ``recipe of learning'' within the
Bayesian approach \cite{Ago03,Dos03}.

Regarding actual practical application of Bayesian inference, the
determination of the prior probability is the most problematic point.
Different recipes for selecting an appropriate prior exist.  Common
choices are distributions which are uniform in the parameters or the
logarithms of the parameters.  Rigorous justifications are in general
not possible and one is left with postulating some heuristic ad hoc
Ansatz.

However, dynamic force spectroscopy usually provides large data sets,
i.e. large $N$.  Then the likelihood (\ref{3.10}) develops a narrow peak
in the region of its maximum $\boldmu^\ast$ (see next section) and the
prior $p(\boldmu,{\bf v})$ in (\ref{3.20}), though usually unknown in
detail, can be considered as approximately constant, i.e. $p({\boldmu}
| {\bf f}, {\bf v}) \propto p({\bf f} | \boldmu, {\bf v})$.  Given
${\bf f}$ and ${\bf v}$, the likelihood (\ref{3.10}) thus quantifies the
``likeliness'' that the ``true'' model parameters are $\boldmu$.

The upshot of the above intuitive considerations is
that maximizing (\ref{3.10}) with respect to $\boldmu$ should yield
the best possible guess for the unknown true parameters.
Furthermore, the
statistical uncertainties of this estimate should be somehow related 
to the width of the likelihood.  
In the following subsection, we leave this intuitive level and 
turn to a more rigorous discussion of the asymptotic properties 
of the maximum likelihood estimator.

\subsection{Asymptotic Properties}
\label{asymptotic}
Let us assume that the rupture forces $f_i$ have been sampled
according to the ``true'' distribution $p_1(f_i|\boldmu_0, v_i)$ with
unknown, {\em ``true'' model parameters} $\boldmu_0$.  For a given set
of rupture forces $\bf{f}$ and pulling velocities $\bf{v}$ the maximum
likelihood estimate can then be determined as described above.  Upon
repeating the entire set of $N$ pulling experiments with the same set
of pulling velocities ${\bf v}$, a different set of rupture data ${\bf
f}$ will be sampled, yielding a different maximum likelihood estimate
$\boldmu^\ast$.  While the probability distribution of ${\bf f}$ is
given by (\ref{3.10}) with $\boldmu=\boldmu_0$, what can we say about
the distribution of the maximum likelihood estimates $\boldmu^\ast$?
 
To answer this question we first exploit the fact that in typical
single-molecule pulling experiments for each pulling velocity several
hundred rupture forces are measured.  The resulting set of rupture
force data $\bf{f}$ is thus quite large and it is convenient to
rewrite the likelihood (\ref{3.10}) as
\begin{eqnarray}
p({\bf f} | \boldmu, {\bf v}) & = & \exp\{-N\,s_N({\bf f}, 
\boldmu , {\bf v})\}
\label{3.30} \\
s_N({\bf f}, \boldmu , {\bf v}) & := & - N^{-1}\sum_{i=1}^N 
\ln p_1(f_i|\boldmu, v_i) \ .
\label{3.40}
\end{eqnarray}
Furthermore, we assume that rupture forces have been measured at $Z$
different pulling velocities $v_\beta$, $\beta=1,...,Z$, 
and that the relative frequency
with which the different pulling velocities $v$ are sampled, converges
towards a well defined limit $\rho(v)=\sum_{\beta=1}^{Z}\rho_\beta
\delta(v-v_\beta )$ for $N\to\infty$.  Then it follows from the law of
large numbers \cite{Cov91} that
\begin{eqnarray}
s_N({\bf f}, \boldmu , {\bf v}) & \to & s(\boldmu):=
- \langle \ln p_1(f|\boldmu,v)\rangle_1
\label{3.50}
\end{eqnarray}
for $N\to\infty$, where $\langle\cdots\rangle_1$ indicates an average
over $f$ and $v$ with weight $p_1(f|\boldmu_0,v)\, \rho(v)$.  Hence,
$s_N$ is an intensive, entropy-like quantity.  Observing that
$s(\boldmu)-s(\boldmu_0)$ is a relative entropy of the form 
$\langle \ln (p_1(f|\boldmu_0,v)/p_1(f|\boldmu,v)\rangle_1$, 
and using the fact
that $p_1(f|\boldmu,v)$ is normalized with respect to $f$
for every choice of the parameters $\boldmu$, we obtain
\begin{eqnarray}
\label{3.60}
s(\boldmu)-s(\boldmu_0) =&&\\ \nonumber
\int dv\;\rho(v)&&\hspace{-0.3cm}
\int df\;p_1(f|\boldmu,v)
\left[ R\ln R - R +1\right]\ ,
\end{eqnarray}
with $R:=p(f|\boldmu_0,v)/p(f|\boldmu,v)$.
Finally, using the inequality
\begin{equation}
0\le\int_1^R dx\;\ln x = R\ln R-R+1 \hspace{0.5cm} \forall R\ ,
\label{3.70}
\end{equation}
we see that $s(\boldmu)-s(\boldmu_0)\geq 0$ and that 
$s(\boldmu)-s(\boldmu_0)= 0$ if the expression in the 
square brackets on the right hand side of (\ref{3.60}) vanishes 
for all $f$.  
Thus, $s(\boldmu)$ has a {\it
unique} absolute minimum at $\boldmu=\boldmu_0$ \cite{Cov91}.  Since
$s_N$ converges for large $N$ toward $s$ according to (\ref{3.50}), also
the minimum $\boldmu^\ast$ of the former converges to the minimum
$\boldmu_0$ of the latter, i.e. the maximum likelihood
estimate is a so-called consistent estimate \cite{Cra46}.

For large, but finite $N$ values, $\boldmu$ will be close to $\boldmu^\ast$.
Consequently, we can expand $s_N({\bf f}, \boldmu , {\bf v})$ up to
second order about its minimum at $\boldmu^\ast$ and neglecting terms
of order ${\mathcal{O}}(1/\sqrt{N})$, the Hessian matrix of $s_N({\bf
f},\boldmu^\ast,{\bf v})$ can be replaced by the Hessian
$H=H(\boldmu_0)$ of $s(\boldmu_0)$, which is generically positive
definite, i.e.
\begin{eqnarray}
s_N({\bf f},\boldmu^\ast+\bolddelta,{\bf v}) & = & s_N({\bf
f},\boldmu^\ast,{\bf v}) + \bolddelta^\dagger H \bolddelta/2 \ .
\label{3.80}
\end{eqnarray}
For large $N$ this is a very good approximation for all
$\boldmu$-values, and $p({\bf f} | \boldmu, {\bf v})$ approaches a very
sharply peaked Gaussian about $\boldmu^\ast$,
\begin{equation}
p({\bf f} | \boldmu, {\bf v}) 
\propto
\exp\{-N(\boldmu-\boldmu^\ast)^\dagger H (\boldmu-\boldmu^\ast)/2\} \ .
\label{3.90}
\end{equation}

We can now determine the first moments of the distribution of the
maximum likelihood estimate $\boldmu^\ast$ (upon many repetitions of
the same experiment).  Differentiating (\ref{3.80}) and choosing
$\bolddelta=\boldmu_0-\boldmu^\ast$, results in
\begin{equation}
\boldmu^\ast-\boldmu_0 = - H^{-1}\partial s_N({\bf f},\boldmu_0,{\bf
  v})/\partial\boldmu \ .
\label{3.100}
\end{equation}
Averaging over ${\bf f}$ yields zero in the right hand side, as can be
inferred from (\ref{3.40}), (\ref{3.50}) and the fact that $\boldmu_0$ is
the minimum of $s$. Hence,
\begin{equation}
\langle\boldmu^\ast\rangle = \boldmu_0 \ ,
\label{3.110}
\end{equation}
where $\langle\cdots\rangle$ indicates an average over $\bf{f}$ with
weight $p(\bf{f}|\boldmu_0,\bf{v})$ for a given set of pulling
velocities $\bf{v}$.  Equation (\ref{3.110}) thus shows that the maximum
likelihood estimate is ``unbiased'' for large $N$.

With (\ref{3.100}) the determination of the second moments is
straightforward.  Using
\begin{eqnarray}
  \left\langle \frac{\partial}{\partial \mu_i} s_N({\bf
  f},\boldmu_0,{\bf v}) \frac{\partial}{\partial \mu_j} s_N({\bf
  f},\boldmu_0,{\bf v}) \right\rangle = \frac{1}{N}H_{ij} \ ,
  \label{3.120}
\end{eqnarray}
gives the covariance matrix for the maximum likelihood estimate:
\begin{equation}
\langle [\boldmu^\ast-\boldmu_0]\, 
[\boldmu^\ast-\boldmu_0]^\dagger\rangle = (N\, H)^{-1} \ .
\label{3.130}
\end{equation}
Observing that $(N\,H)^{-1}$ is also the covariance matrix of the
distribution from (\ref{3.90}) we arrive at our

{\em First main conclusion}: For any given, sufficiently large data
set ${\bf f}$, the expected deviation of the concomitant maximum
likelihood estimate $\boldmu^\ast$ from the ``true'' parameters
$\boldmu_0$ immediately follows from the ``peak-width'' of likelihood
(\ref{3.10}), considered as a function of $\boldmu$.

Similarly, using the central limit theorem, one can show (see Appendix
\ref{gaussian_dist}), that $\boldmu^\ast$ is Gaussian distributed,
yielding with (\ref{3.90}) our

{\em Second main conclusion}: Apart from the peak position and a
normalization factor, the likelihood (\ref{3.10}) for one given data set
${\bf f}$ looks practically the same as the distribution of the
maximum likelihood estimates $\boldmu^\ast$ from many repetitions of
the $N$ pulling experiments.

\subsection{Cram\'er-Rao bound}
\label{cramer}
It should be noted that, in order to derive the above two main
conclusions, we did not make any use of the Bayesian formalism
(\ref{3.20}) at all.  The latter only served to acquire an intuitive
idea about the meaning of the likelihood (\ref{3.10}).  At this
intuitive level, we have seen that the left hand side of (\ref{3.20}) is
very well approximated by the sharply peaked Gaussian in (\ref{3.90})
and hence it is reasonable to expect that its maximum $\boldmu^\ast$
should be the best possible guess for the unknown true parameters
$\boldmu_0$ that possibly can be inferred from a given set of data
${\bf f}$.  A more rigorous line of reasoning starts with an arbitrary
``recipe'' $\tilde{\boldmu}({\bf f})$ of estimating the true parameters
$\boldmu_0$ from a given data set ${\bf f}$.  The only assumption is
that this recipe is unbiased, i.e. upon repeating the same experiment
many times, the ``true'' parameters are recovered on average,
$\langle \tilde{\boldmu} ({\bf f})\rangle =\boldmu_0$.  By
generalizing the well-known Cram\'er-Rao inequality \cite{Cov91},
which in turn is basically a descendant of the Cauchy-Schwarz
inequality, one can show \cite{Cra46} for any such ``recipe''
$\tilde{\boldmu} ({\bf f})$ that
\begin{equation}
\langle [\tilde{\boldmu}-\boldmu_0]\, 
[\tilde{\boldmu}-\boldmu_0]^\dagger\rangle - (N\, H)^{-1} \geq 0 \ ,
\label{3.140}
\end{equation}
i.e. the matrix in the left hand side is non-negative definite.
Comparison with (\ref{3.130}) yields our

{\em Third main conclusion:} There is no unbiased estimator
$\tilde{\boldmu}$ of the true parameters $\boldmu_0$ which on the
average outperforms the maximum likelihood estimate
$\boldmu^\ast$.

The remaining possibility that a biased estimator may be even better
is rather subtle to treat rigorously, but intuitively this seems quite
unlikely.  Furthermore, in the above conclusion we exploited the
relation (\ref{3.130}) which is strictly correct only for asymptotically
large $N$.  Finally, also the criterion of minimizing the left hand
side in (\ref{3.140}) itself is in principle debatable, but hardly in
practice.

\subsection{Parameter Inference: Main steps for the practical application}
\label{steps}
We now briefly summarize the main steps of 
the maximum likelihood method for evaluating
single-molecule pulling experiments. 
The first step consists in specifying the dependence of the rupture 
force probability $p_1(f|\boldmu, v)$ on the model parameters 
$\boldmu$ and the pulling velocity $v$ within any given 
theoretical description.
Then for the set of rupture
forces ${\bf f} = \{f_i\}_{i=1}^N$ and corresponding pulling
velocities ${\bf v} = \{v_i\}_{i=1}^N$ the logarithm of the likelihood
(\ref{3.30})
\begin{equation}
\label{3.150}
-N s_N({\bf f}, \boldmu , {\bf v})  =  \sum_{i=1}^N 
\ln p_1(f_i|\boldmu, v_i)
\end{equation}
is maximized with respect to the model parameters $\boldmu$.  Usually
this step has do be accomplished numerically.  The position of the
maximum defines the most probable parameters $\boldmu^\ast$ which are
on average closer to the true model parameters than any other
estimate.  The statistical uncertainties of the parameters can then be
estimated as
\begin{equation}
\langle [\boldmu^\ast-\boldmu_0]\, 
[\boldmu^\ast-\boldmu_0]^\dagger\rangle \approx (N\, H_N)^{-1} 
\ ,
\label{3.160}
\end{equation}
where $H_N$ denotes the Hessian matrix of $s_N$ evaluated at the most
probable parameters $\boldmu^\ast$ and $\boldmu_0$ the true model
parameters.  Moreover, the distribution of the maximum likelihood
estimate is Gaussian with mean $\boldmu_0$.

\section{Application to single-molecule force spectroscopy: Bell's model}
\label{s2a}
Combining the result (\ref{2.20}) with approximations (\ref{2.30}),
(\ref{2.40}) [or (\ref{2.50}), (\ref{2.60})], one can apply the
maximum likelihood approach from the previous section to deduce the
rate parameters for any exponentially increasing escape rate
(\ref{1.20}). In what follows, however, we will focus on two efficient
rate approximations [see Eqs.~(\ref{4.10}) and (\ref{5.10}) below],
which allow one to evaluate the integral from Eq.~(\ref{2.21})
analytically.

The first common approximation, originally due to Bell \cite{Bel78},
consist in the linearization of the force dependent 
potential barrier according to (\ref{1.30})  and in
neglecting the force dependence of the pre-exponential 
factor $\omega(f)$ in the Kramers rate (\ref{1.20}), 
resulting in [cf. (\ref{1.30}) and Fig. 2]
\begin{eqnarray}
k(f) = k_0\exp\left(\frac{\xb f}{\kBT}\right) =: \exp(\lambda +
\alpha f) \ ,
\label{4.10}
\end{eqnarray}    
where $k_0:=\omega(0)\exp(- \Eb(0)/\kBT)$ is the
force-free dissociation rate, $\xb $ the dissociation length (distance
between potential well and barrier), and $\lambda:=\ln k_0$, $\alpha
:= \xb /\kBT$ are convenient abbreviations.

Substituting Eq.~(\ref{4.10}) into (\ref{2.20}), (\ref{2.21}) 
and going over to $f$
as independent variable, a straightforward calculation yields the
probability density of rupture events for Bell's model:
\begin{equation}
  p_1(f|\boldmu, \ v) = 
  \frac{e^{\lambda + \alpha f}}{\kappa v}
  \exp\left(-\frac{e^\lambda}{\kappa v}\frac{e^{\alpha f}-1}{\alpha}\right)
  \ .
  \label{4.20}
\end{equation}
The rupture force density (\ref{4.20}) is conditioned on 
$\boldmu = (\lambda, \alpha)$, and $v$.
As usual, we assume that the pulling velocity $v$ is known
exactly for each measurement, and similarly for the 
elasticity $\kappa$ appearing on the right hand side of (\ref{4.20}). 
The remaining model parameters to be estimated
from a given set of rupture forces ${\bf f} = \{f_i\}_{i=1}^N$ 
measured at pulling velocities $\{v_\beta\}_{\beta=1}^Z$ 
with relative frequencies $\rho_\beta$ are therefore 
$\boldmu = (\lambda, \alpha)$.
%
%

\subsection{Statistical uncertainties of Bell's model}
\label{s4a}

For the above specified model we can calculate $s(\boldmu) = -\langle \ln
p_1(f|\boldmu, v) \rangle_1$ as defined in (\ref{3.50}) analytically, 
if the dimensionless quantity $\tau_\beta := e^{\lambda_0}/(\kappa v_\beta
\alpha_0)$ is small for all pulling velocities.  
In fact, $\tau_\beta<1$ is equivalent to the assumption that the distribution
of rupture events $p_1(f|\boldmu_0, v_\beta)$ has a maximum at some
force $f^\ast_\beta >0$.  Details of the calculations are given in
Appendix \ref{sigma_bell}, resulting in:
\begin{eqnarray}
\label{4.30}
s(\boldmu) &=& -\lambda + \sum_{\beta=1}^{Z} 
\rho_\beta s_\beta(\boldmu)
+ {\mathcal{O}}(\tau_\beta)\ ,\\
\nonumber
s_\beta(\boldmu) &:=& \ln(\kappa v_\beta)
+\eta({\mathcal{C}} +\ln \tau_\beta)
+\frac{e^\lambda\, \Gamma\left(\eta+1\right)}{\kappa v_\beta \alpha\, [\tau_\beta]^{\eta}}
\end{eqnarray} 
with $\eta:=\alpha / \alpha_0$, ${\mathcal{C}} \approx 0.577$ Euler's
constant, 
and $\Gamma(\cdot)$ the Gamma function.  Differentiating
(\ref{4.30}) twice with respect to the model parameters, a
straightforward calculation yields the Hessian $H = H(\boldmu_0)$ of
$s(\boldmu_0)$.  Finally, inverting this ($2\times 2$) matrix, we
obtain the variance of the maximum likelihood estimate of the two
parameters:
\begin{eqnarray}
  \label{4.40}
\!\!\!\!\!\!\!\!\!\!\!\!  &&\langle (\alpha^\ast - \alpha_0)^2 \rangle 
  \approx
  \frac{\alpha_0^2}{N}
  \frac{1}{\frac{\pi^2}{6}+\sigma^2(\ln(r))}\ ,
  \\
\!\!\!\!\!\!\!\!\!\!\!\!  &&\langle (\lambda^\ast - \lambda_0)^2 \rangle 
  \approx
  \frac{1}{N}
  \frac{\sum\limits_{\beta=1}^Z \frac{N_\beta}{N}\left(
    \alpha_0^2\left<f_\beta\right>^2
    +\frac{\pi^2}{6}
    \right)}
       {
         \frac{\pi^2}{6}+\sigma^2(\ln(r))
       } \ ,
  \label{4.50}
\end{eqnarray}
where
\begin{equation}
\sigma^2(\ln(r)) = \sum_{\beta=1}^Z \rho_\beta \ln^2(r_\beta) -
\left(\sum_{\beta=1}^Z \rho_\beta \ln(r_\beta)\right)^2
\label{4.55}
\end{equation}
is the
variance of the logarithm of the loading rate $r:=\kappa v$ and
\begin{equation}
\left<f_\beta\right> = -1/{\alpha_0} ({\mathcal C} +
\ln(e^{\lambda_0}/(r_\beta\alpha_0)))
\label{4.57}
\end{equation}
the expected rupture force at
loading rate $r_\beta$ (again neglecting terms of order ${\mathcal
O}(\tau_\beta)$).  For practical application of Eqs.  (\ref{4.40}) and
(\ref{4.50}) the true model parameters $\boldmu_0$ in the right-hand
sides of the equations have to be replaced by the concomitant maximum
likelihood estimate $\boldmu^\ast$.  For large $N$ this is a very good
approximation.

In single-molecule pulling experiments only a limited range of pulling
velocities $v$ is accessible, i.e.  $v_\beta \in [\vmin , \vmax]$.
Now the question arises: for which distribution of pulling velocities
$\rho(v)$ do the statistical uncertainties (\ref{4.40}), (\ref{4.50}) of the
estimated parameters become minimal?  Recognizing that the variance of
the maximum likelihood estimate of $\alpha$ depends on the
distribution of the pulling velocities solely via the term
$\sigma^2(\ln(r))$ in the denominator of (\ref{4.40}), while this
distribution enters into the expression for the statistical
uncertainties of $\lambda^\ast$ also via the terms
$\left<f_\beta\right>$ in the numerator of (\ref{4.50}), we see, that
it is, in general, not possible to simultaneously minimize the two
uncertainties.  Given a fixed number $N$ of pulling experiments, the
minimization of the variance $\langle (\alpha^\ast - \alpha_0)^2
\rangle$ of $\alpha^\ast$ is equivalent to a maximization of
$\sigma^2(\ln(r))$.  Under the constraint $v_\beta \in [\vmin,
\vmax]$ for all pulling velocities $v_\beta$, this maximum is
obviously reached if half of the rupture forces have been sampled at a
pulling velocity as large as possible, i.e. $\vmax$, and the other
half at a pulling velocity as small as possible, i.e. $\vmin$.  This
result is independent of the values of the true model parameters
$\boldmu_0$.  Regarding the distribution of $\lambda^\ast$, the
situation is more complicated.  Given the analytic expression
(\ref{4.50}) for the error, the calculation is straightforward, but not
shown here for the following two reasons. The ''best'' choice consists
again in sampling just at the two extreme pulling velocities $\vmin$
and $\vmax$.  The relative number of pulling experiments for each of
the two pulling velocities is, however, non-trivial and depends on the
values of the true model parameters $\boldmu_0$.  It is, therefore (in
contrast to the result for $\alpha^\ast$), only of limited use for a
real experiment.

\subsection{Illustration for computer generated data}
\label{s4d}
\begin{figure}
{\leftline {(a)}}
\includegraphics[width=0.85\linewidth]{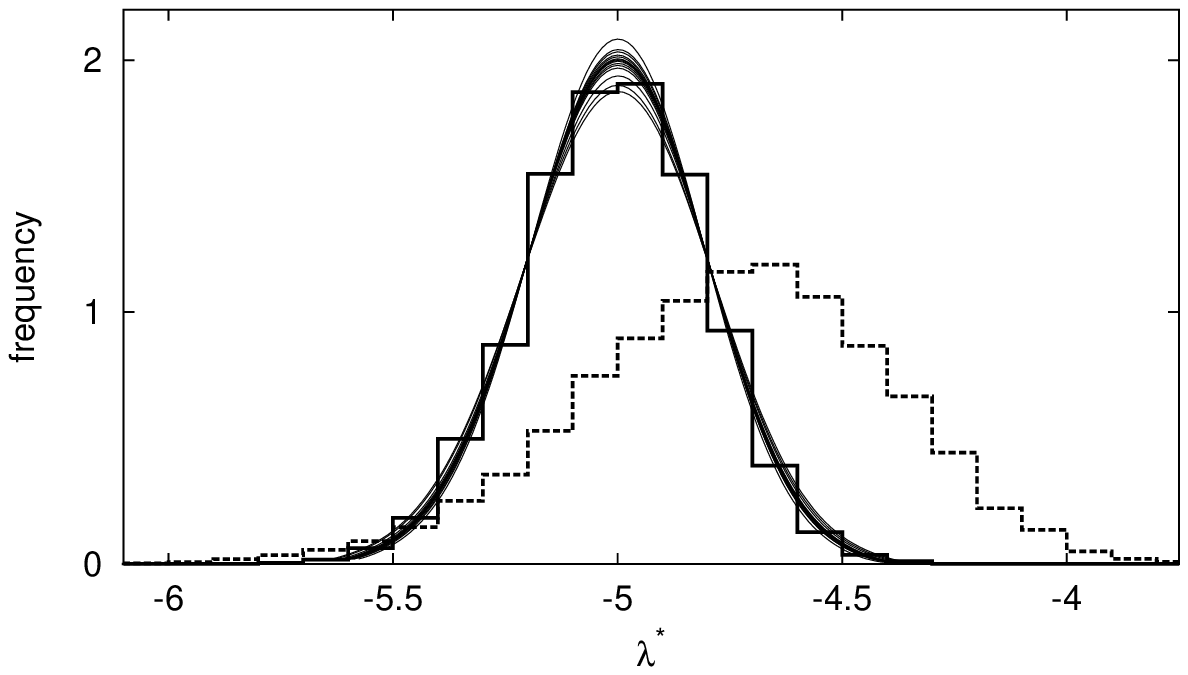}
{\leftline {(b)}}
\includegraphics[width=0.85\linewidth]{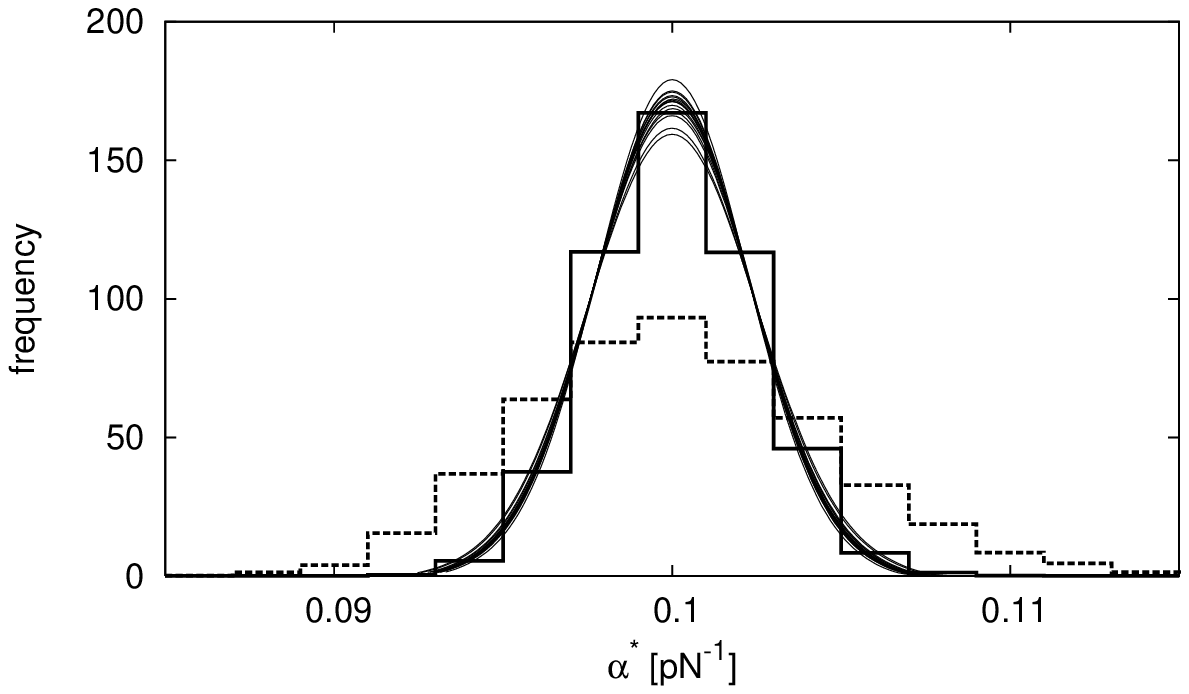}
\caption{Solid histogram: Distributions of the first and second 
components of the maxima $\boldmu^\ast=(\lambda^\ast,\alpha^\ast)$ of
the likelihood (\ref{3.10}) for 10000 ``computer experiments''.  For
each of them, $N=400$ rupture forces $f$ were sampled according to
(\ref{4.20}), 100 for each of the $4$ loading rates $\kappa v=50$, $200$,
$1000$, $5000$ pN/s and with ``true'' parameters $\lambda_0=-5$ and
$\alpha_0=0.1$ pN$^{-1}$.  These are typical numbers in ``real
experiments'' \cite{Mer01}.  For sake of better visibility the
bin-width of the histograms is much larger than the optimal bin width
for a Gaussian distribution (Appendix \ref{opt_bin}).  Thin lines:
Likelihood (\ref{3.10}) for the first 15 of the 10000 experiments after
integrating over the other component, shifting the maximum to
$\mu_{i,0}$, and normalizing (some are almost indistinguishable).
Dotted histogram: Distribution of the estimates for $\lambda$ and
$\alpha$ according to the ``standard method'', as described in the
main text. }
\label{fig3}
\end{figure}

We now illustrate the findings from sections \ref{asymptotic} and
\ref{cramer}.  To avoid uncontrollable experimental inaccuracies and
uncertainties regarding the ``true'' model and the ``true'' model
parameters $\boldmu_0$, we numerically generated synthetic rupture
data ${\bf f}$ by ``simulating'' an idealized experiment on the
computer according to the probabilistic ``laws'' (\ref{4.10}), (\ref{4.20})
with given parameters $\boldmu=\boldmu_0$.  Hence all remaining
uncertainties are statistical finite $N$ effects.  Fig. \ref{fig3}
shows the results for 10000 repetitions of a computer experiment, each
sampling $N=400$ rupture forces ${\bf f}$ according to (\ref{4.20}) with
experimentally realistic parameter values $\alpha_0 = 0.1\,$pN$^{-1}$
and $\lambda_0 = -5$.  Since two-dimensional distributions are
difficult to compare graphically, we focus on the marginal
distributions.  For each of the 10000 experiments, the maximum
$\boldmu^\ast=(\lambda^\ast,\alpha^\ast)$ of the likelihood
(\ref{3.10}), considered as a function of $\boldmu$, was determined
numerically.  The distribution of the resulting $\lambda^\ast$- and
$\alpha^\ast$-values are depicted as histograms in Fig. \ref{fig3}.
The standard deviations of the maximum likelihood estimate for
$\lambda$ and $\alpha$, determined from the 10000 experiments are
$s(\lambda^\ast )=0.20$ and $s(\alpha^\ast ) =0.0023\,$pN$^{-1}$
respectively.  These values coincide with those obtained from the
analytical approximations (\ref{4.40})-(\ref{4.57}) up to the third
non-vanishing digit.  Replacing the parameters $\boldmu_0$ on the
right-hand side of (\ref{4.40}), (\ref{4.50}) by the maximum likelihood
estimate $\boldmu^\ast$ for one given data set thus provides reliable
estimates for the statistical uncertainties.

Furthermore, for the first 15 of the 10000 experiments, after
integrating over the other parameter in the likelihood (\ref{3.10}),
shifting the peak position from $\mu_i^\ast$ to $\mu_{i,0}$, and
normalizing, the resulting marginal distributions were plotted in
Fig. \ref{fig3}.  They closely agree with the histograms.  These
observations illustrate very convincingly our two conclusions from
section \ref{asymptotic} above.  In particular, finite-$N$ corrections
are apparently very small for the typical parameter values used in
this example.  Also the practically perfect Gaussianity of the
distributions is as expected, cf. (\ref{3.90}).

Let us finally compare the performance of the maximum likelihood
estimate with that of the most widely used ``recipe'' of parameter
estimation in the field of single-molecule pulling experiments.  This
consist of the following steps: (i) Fit a Gaussian to the observed
rupture force distribution for a fixed pulling velocity $v$ and
approximate the most probable rupture force $f^\ast$ by the maximum of
that Gaussian.  (ii) Plot $f^\ast$ for different $v$ versus $\ln (v)$
and fit the resulting points by a straight line.  (iii) Assume that
the model (\ref{4.10}), (\ref{4.20}) is applicable and deduce its model
parameters $\boldmu=(\lambda,\alpha)$ from the slope and the axis
intercept of the straight line as detailed e.g. in
\cite{Eva97,Mer01,Mer99,Str00,Evs03}.  We have applied this procedure
to each of the 10000 experiments in Fig. \ref{fig3} and plotted the
distribution of the resulting estimates for $\lambda$ and $\alpha$ in
Fig. \ref{fig3}.  The systematic bias of the estimate for $\lambda$
can be traced back to fitting a Gaussian, which is symmetric about its
maximum, to an asymmetric ``true'' distribution (\ref{4.20})
\cite{Evs03}, while the suboptimal variance of the estimate for both
$\lambda$ and $\alpha$ signals that quite some information is lost by
only considering the most probable rupture forces $f^\ast$.  Hence,
the maximum likelihood estimate represents a substantial improvement
compared to the so far ``standard method'' of data evaluation in this
field.  This is in agreement with our conclusion from section
\ref{cramer}.  We have also directly compared the maximum likelihood
estimate with other known ``recipes'' of evaluating single-molecule
rupture data, e.g. \cite{Evs03}.  In all cases we found that the
maximum likelihood was superior.

\section{Extension of Bell's model}
\subsection{Rate Ansatz}
\label{s2b}
As can be seen from Fig.~\ref{fig2}, a linearization (\ref{4.10})
of the force dependent potential barrier $\Eb(f)$ is
relatively good for small forces. 
For a larger forces, the distance between potential extrema decreases, 
leading to a weaker sensitivity of the barrier height to force 
variations upon further pulling than in Eq.~(\ref{4.10}).

Models including this effect, in general, rely on some assumptions
concerning the shape of the energy landscape.  Typical choices are
Morse potentials, harmonic potentials with a cusp barrier, two
parabolas at the potential extrema joined at a midpoint, and linear-cubic
potentials \cite{Dud03, Hum03, Han06, Mal06, Dud06, Han06, Hus08,
Rie03}.
It has been suggested in \cite{Dud06} that for sufficiently high
barriers, i.e. forces substantially smaller than the critical force,
the dissociation rate can be written in a unique form
\begin{equation}
k(f)=(1-\gamma\alpha f/\epsilon)^{1/\gamma-1}\, 
e^{\lambda+\epsilon [1-(1-\gamma\alpha f/\epsilon)^{1/\gamma}]}
\label{5.10}
\end{equation}
with three model parameters ${\boldmu}=(\lambda,\alpha,\epsilon)$ and
fixed exponent $1/\gamma$.  Here, $\lambda$ and $\alpha$ have the same
physical meaning as in Eq. (\ref{4.10}), and $\epsilon:=\Eb(0)/\kBT$
stands for the force-free activation energy barrier in units of the
thermal energy $\kBT$. 

The extra parameter $\gamma$ controls the manner in which the barrier
height decreases with the applied force. 
We note that, physically,
this parameter should be in the range $\gamma \in (0, 1]$ since
$\gamma\leq 0$ would imply a positive barrier for all $f>0$. 
On the other hand, the first derivative of the barrier height equals minus
the distance between the potential extrema corresponding to a given
force value. Since we expect this distance to decrease with the force,
we conclude that the second force derivative of the barrier height
must be positive, excluding $\gamma$-values greater than 1. 
Specifically, for $\gamma=1$ the parameter $\epsilon$ drops out and 
one recovers Bell's model (\ref{4.10}), $\gamma=2/3$ reproduces the
Kramers rate for a cubic reaction potential, and $\gamma=1/2$
corresponds to a parabolic potential well with a cusp barrier.

Substituting Eqs.~(\ref{2.10}), (\ref{5.10}) into (\ref{1.10}), one
derives the survival probability of the bond up to force $f$:
\begin{equation}
  n(f) = \exp\left( -\frac{e^{\lambda}}{\kappa v} \frac{e^{\epsilon
  [1-(1-\gamma\alpha f/\epsilon)^{1/\gamma}]} -1}{\alpha} \right) \ ,
  \label{5.20}
\end{equation}
and the probability density of rupture events follows from
Eq.~(\ref{2.20}).

As already mentioned, the application of Kramers reaction rate theory
requires that the potential barrier $\Eb(f)$ be sufficiently high
(compared to the thermal energy $\kBT$).  Thus, all approximations are
only valid for forces substantially smaller than the critical force at
which the barrier vanishes, $f_\mathrm{c} = \epsilon/(\gamma \alpha)$.

The above discussion suggests that dynamic force spectroscopy should, in
principle, provide the possibility, not only to determine the
force-free dissociation rate $k_0 = \exp(\lambda)$ and the dissociation
length $\xb$, but also the
force-free activation energy barrier $\Eb(0)=\epsilon \kBT$.
Naturally, the question arises how accurate these estimates will be
and whether the inferred values critically depend on the chosen
theoretical model, in particular on the value of the parameter
$\gamma$.  These questions will be addressed next.

\subsection{Numerical experiment}
\label{s4b}

\begin{figure}
\includegraphics[width=0.95\linewidth]{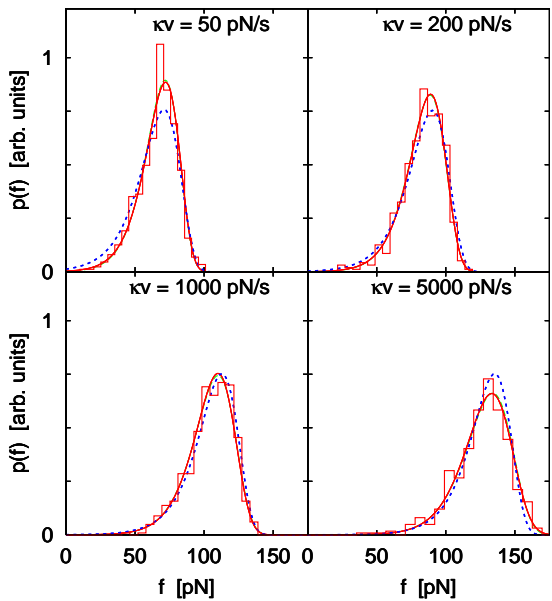}
\caption{
(Color online)
Rupture force distribution for different loading rates $\kappa v$.
Histograms: numerically generated rupture forces according to
(\ref{2.20}), (\ref{5.10}) with $\gamma=2/3$, $\lambda_0=-5$, $\alpha_0=0.1$
pN$^{-1}$, $\epsilon_0=15$.  For each $\kappa v$, we sampled 500
forces, i.e. $N=2000$.  The bin-width is chosen according to
eq. (\ref{A15}).  The maximum likelihood fits $p_1(f|\boldmu^\ast,v)$
according to (\ref{2.20}), (\ref{5.10}) for $\gamma=1/2$ and $\gamma=2/3$
(red solid) are not distinguishable within the line width in this
plot.  For $\gamma = 1/2$, the fit parameters have the following
values: $\lambda = -5.37$, $\alpha = 0.110$, and $\epsilon =
17.2$. For $\gamma = 2/3$, the fit results are slightly closer to the
true parameter values, namely, $\lambda = -5.20$, $\alpha = 0.104$, and
$\epsilon = 14.64$. Blue dashed distribution: same for Bell's Ansatz
$\gamma=1$, with fit results $\lambda = -3.81$ and $\alpha = 0.072$.
Upon repeating the entire ``numerical experiment'', the resulting
plots always look practically the same.
}
\label{fig4}
\end{figure}

\begin{figure*}
\includegraphics[width=2\columnwidth]{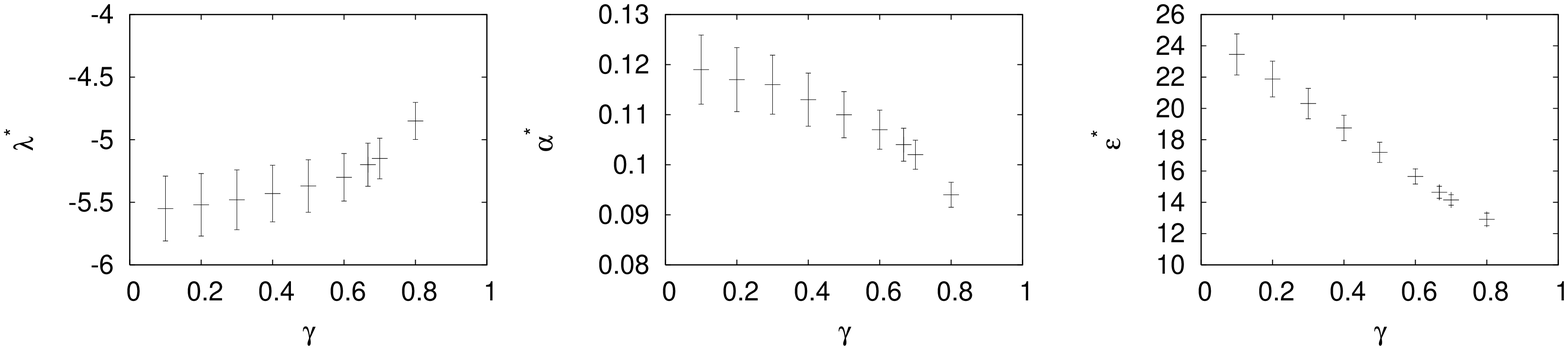}
\caption{Maximum likelihood fit values for the data set from
  Fig.~\ref{fig4} fitted with rupture force distribution (\ref{2.20})
  supplemented with the escape rate (\ref{5.20}). Each data point was
  obtained by fitting the same data set, but assuming a different
  value of the parameter $\gamma$.}
\label{fig5}
\end{figure*}

\begin{figure*}
\includegraphics[width=2\columnwidth]{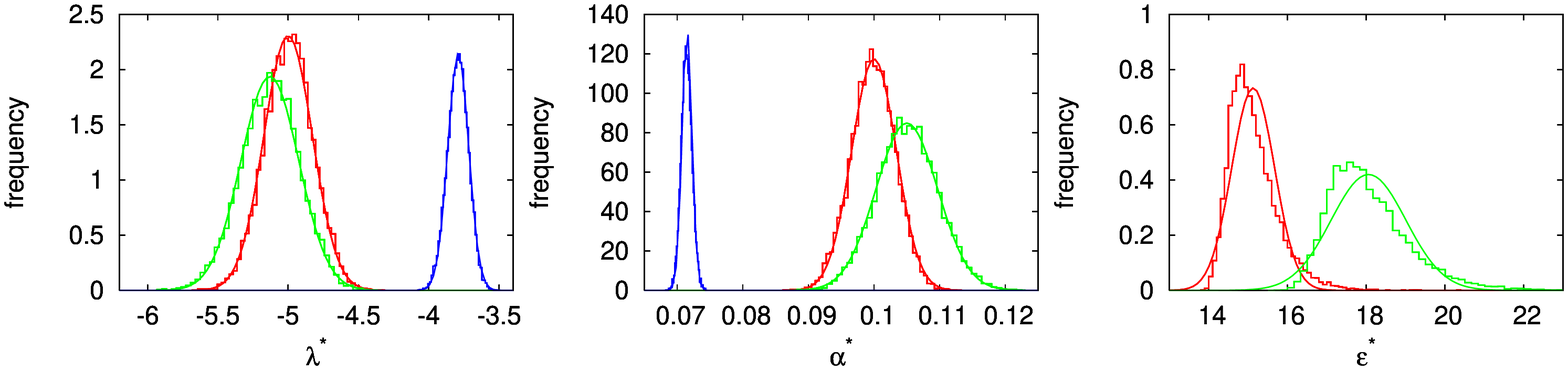}
\caption{
(Color online)
Histograms: Numerically determined distribution of the single
components of the maxima $\boldmu^\ast = (\lambda^\ast, \alpha^\ast,
\epsilon^\ast)$ of the likelihood (\ref{3.10}), (\ref{5.10}) with
$\gamma=1/2$ (green), $\gamma=2/3$ (red) and $\gamma=1$ (blue) for
10000 ``computer experiments''.  For all data sets the rupture forces
were generated numerically according to (\ref{2.20}), (\ref{5.10}) with
$\gamma=2/3$, $\lambda_0=-5$, $\alpha_0=0.1$ pN$^{-1}$,
$\epsilon_0=15$.  For each $\kappa v$, we sampled 500 forces,
i.e. $N=2000$.  The bin-width of the histograms is $h_\beta = 3.49
s_{N_\beta} N_\beta^{-1/3}$ (see Appendix \ref{opt_bin}).  Solid
lines: Gaussian approximations to the distributions with covariance
matrices (\ref{3.130}).  They have been shifted so that their
maximum coincides with the mean of the empirically determined
distributions.  For sake of better visibility the distributions for
$\gamma=1$ have been rescaled by an appropriate value.
}
\label{fig6}
\end{figure*}

There is an ongoing debate in
literature about which of the three exponents, i.e. $\gamma=1$,
$\gamma=2/3$, or $\gamma=1/2$, is most appropriate to use when
evaluating experimental rupture data
\cite{Dud03,Hum03,Mal06,Dud06,Han06}.  Taking for granted that one of
the three models approximates the ``truth'' satisfactorily, choosing
$\boldmu=\boldmu^\ast$ is -- according to our above conclusions -- the
closest one can get to the ``full truth'' on the basis of one given
data set ${\bf f}$.  In case of disagreement about the ``true''
$\gamma$-value, a fully objective selection criterion seems unavailable 
in principle.  
In practice, the usual criterion is the
comparison with the basic ``true'' quantity observed experimentally,
namely the distribution of rupture forces.

For the example shown in figure \ref{fig4} we sampled $N=2000$ rupture
forces ${\bf f}$ according to the distribution (\ref{2.20}) with rate
(\ref{5.10}). In the simulations, we have taken $\gamma=2/3$ and
realistic parameter values $\lambda_0=-5$, $\alpha_0=0.1$ pN$^{-1}$,
$\epsilon_0=15$. 

In order to test the maximum likelihood method, for the resulting data
set we determined the estimate $\boldmu^\ast$ for several possible
$\gamma$-values. Since in a real experiment, the value of the exponent
$\gamma$ is not known a priori, also during the fitting, the
$\gamma$-values used were not necessarily coincident with the ``true''
value used for data generation. The bin-width of the histograms in
Fig. \ref{fig4} was chosen as $h_\beta = 2.83 s_{N_\beta}
N_\beta^{-1/3}$, where $s_{N_\beta}$ is the standard deviation of the
rupture forces measured at pulling velocity $v_\beta$.  As discussed
in Appendix \ref{opt_bin}, this is the optimal choice of the bin-width
for the Bell model; although this model is in fact not the true one in
our numerical experiment, this choice of the bin-width remains
suitable, as every distribution separately can be very well fitted
with Eq.~(\ref{4.20}).

Comparing in Fig.~\ref{fig4} the resulting distributions
$p_1(f|\boldmu^\ast, v)$ for three different $\gamma$ values, we
observe the following.  Already the two-parametric Bell's Ansatz
($\gamma = 1$) reproduces the experimental distribution of rupture
forces with reasonable accuracy, see the dotted line in
Fig.~\ref{fig4}. However, if one increases the number of fit
parameters to three, one obtains the fit distributions notably
different from the Bell's curves, see the solid lines corresponding to
$\gamma = 1/2$ and $\gamma = 2/3$. We also note that the difference
between the curves corresponding to these two values of $\gamma$ is
smaller than the line thickness. This means that if one treats the
exponent $\gamma$ as a fourth fit parameter, then its precise value
cannot be determined by fitting the experimental rupture force
distribution. With respect to the three remaining fit parameters,
their values are rather close to each other for the fits with $\gamma
= 1/2$ and $\gamma = 2/3$, with the largest discrepancy between the
fit values of the force-free barrier height (see the caption in
Fig.~\ref{fig4}).

When fitting the real experimental data, one does not know a priori
the true value of the exponent $\gamma$. It is therefore of interest
to find out, how the remaining fit parameter values depend on the
assumption with respect to this quantity. Fig.~\ref{fig5} shows the
fit results obtained for different assumed values of $\gamma$ 
within the physically meaningful range. All of the
fitting curves obtained for different $\gamma$-values from
Fig.~\ref{fig5} coincided within the line thickness. We observe that
the resulting fit values of the force-free escape rate $e^\lambda$ and
the dissociation length $\kBT \alpha$ are not very sensitive to the
choice of the exponent $\gamma$. At the same time, the force-free
barrier height value inferred from the fit depends approximately
linearly on the choice of this parameter, and can assume values
differing by as much as a factor of 2 at extreme $\gamma$-values. This
means that when fitting the experimental data, the value of the
force-free barrier height will be determined with the least accuracy. 

In view of these observations, an interesting question arises: are the
approximations (\ref{3.130}) and (\ref{4.40})-(\ref{4.57}) for the
statistical uncertainties of the model parameters still valid for
those models, about which we (in our case) {\it know} that they are
not true? To study this point, we have repeated the above described
procedure for 10000 data sets, each generated in the same way and with
same ``true'' parameters as the data set shown in Fig.~\ref{fig4}.
The distributions of the inferred parameters $\boldmu^\ast$ for the
three different $\gamma$-values are depicted in Fig.~\ref{fig6} as
histograms.  Following section \ref{asymptotic}, they should be
bell-shaped with variance given by Eq.~(\ref{3.130}).  We have
evaluated this expression at the mean values of the inferred
parameters $\boldmu^\ast$ for each of the three $\gamma$-values
respectively.  For $\gamma=2/3$ and $\gamma=1/2$ this had to be done
numerically, whereas for $\gamma=1$, Eqs.~ (\ref{4.40})-(\ref{4.57}) 
could be employed.  The resulting distributions, centered
about the mean value of the inferred parameters, are shown in
Fig.~\ref{fig6} as solid lines.  For $\lambda$ and $\alpha$ they
closely agree to the histograms.  With respect to the parameter
$\epsilon$, the approximated variance agrees very well with the
empirically determined, but finite-$N$ corrections to the full
distributions which are not symmetric about their center are apparent.

In conclusion, although the models for $\gamma=1$ and $\gamma=1/2$ are
not the true models, equation (\ref{3.130}) still yields very good
approximations for the statistical uncertainties of the parameters,
i.e.  for their distribution upon repeating the same experiment many
times.  Nevertheless, comparing these distributions to the
distribution of the parameters for the ``true'' gamma-value $\gamma=2/3$,
we see, that by choosing the wrong model, the systematic deviations to
the parameters are much larger than the statistical uncertainties.  As
one will, in general, be unsure about the true underlying energy
landscape, and thus about the ``true'' $\gamma$-value, this point is
essential if one wishes to use the inferred parameter values in
another context than the interpretation of single-molecule pulling
experiments.

We would like to mention that the deviations between the distributions
resulting from the different $\gamma$ values increase with the pulling
velocity \cite{Dud06}.  Hence, by increasing the range of accessible
pulling velocities $[\vmin, \vmax]$ a clearer distinction between
the models is possible.  However, for precise measurements with the
AFM the loading rate $\kappa v$ is limited to a few orders of
magnitude, comparable with our values.

\section{Conclusions}
\label{conclusion}
In this work we have shown that the maximum likelihood approach 
is an extremely simple, general, and powerful method for parameter 
estimation in the contect of single-molecule force spectroscopy.  
For large data sets it outperforms all other estimates.  
Furthermore, approximations to the
statistical uncertainties of the parameters are available once the
parameters are estimated.  In the case of the standard Bell model we
were able to derive an analytical expression for these uncertainties
in terms of the model parameters and the distribution of the applied
loading rates.  For more general models, the uncertainties can be
determined numerically.

When fitting the experimental data, one usually adopts some functional
form of the force-dependent escape rate involving several fit
parameters.  By means of a numerical example, we have demonstrated
that the largest number of such parameters that can be determined from
the experiment is three. These parameters are related to the
force-free value of the rate, the dissociation length,
and the barrier height in the absence of the force. Furthermore, when
fitting the experimental rupture force distributions, one needs to
make an additional assumption about the manner in which the escape
rate decreases with the applied force. While the fit values of the
force-free escape rate and the dissociation length depend only weakly
on this assumption, the value of the force-free barrier height can be
determined much less reliably.  We have shown that even if the
model adopted for the description of the experiment is not the true
one but predicts distributions of rupture forces similar to the
measured distribution, the statistical uncertainties found from the
maximum likelihood method very well approximate the dispersion of the
estimated parameters upon repeating the same experiment many times.
Often these uncertainties are much smaller than the systematic error
resulting from choosing the ``wrong'' model.

\acknowledgments
We are grateful to the Deutsche Forschungsgemeinschaft 
(RE 1344/3-1, and SFB 613) for financial support of this work.

\begin{appendix}

\section{Asymptotic distribution of the maximum
likelihood estimate}
\label{gaussian_dist}
In this section we proof the second main conclusion from section
\ref{asymptotic}, namely that the distribution of the maximum
likelihood estimate $\boldmu^\ast$ is Gaussian and looks, apart from
the peak position, the same as the likelihood (\ref{3.90}) for one given
data set $\bf{f}$.

To keep things as simple as possible, we assume that there exist
(small) integers $n_\beta$ for each of the $Z$ pulling velocities
$v_\beta$ so that $n_\beta/n_\gamma = \rho_\beta/\rho_\gamma$.
Denoting by $n$ the sum of all $n_\beta$, the total number of rupture
forces can be written as $N=N'n$ and the set of $N$ rupture forces
$\bf{f}$ can be divided into $N'$ subsets ${\bf
f}^k=\{f_{i,\beta}^k\}$ where for fixed $k$, $f_{i,\beta}^k$ is one
out of the $n_\beta$ rupture forces sampled at pulling velocity
$v_\beta$.

Having introduced this notation, we define $N'$ new random variables
\begin{equation}
{\bf X}^k={\bf X}^k({\bf f}^k) = \frac{1}{n}
\sum_{\beta = 1}^{Z}\sum_{i=1}^{n_\beta}
H^{-1} \partial/\partial \boldmu 
\ln p_1(f_{i,\beta}^k|\boldmu_0,v_\beta)
\ .
\label{A0}
\end{equation}
We know already from our discussion in section \ref{asymptotic} that
these random variables have an expectation value zero and a covariance
matrix
\begin{equation}
\langle {\bf X\ X}^\dagger \rangle = \frac{1}{n}H^{-1}
\ .
\label{A1}
\end{equation}
Then, from the central limit theorem \cite{Cra46} it follows that
\begin{equation}
\boldmu^\ast - \boldmu_0 
=\frac{1}{N'}\sum_{k=1}^{N'}{\bf X}^k
\label{A2}
\end{equation}
is Gaussian distributed with mean zero and covariance matrix
\begin{equation}
\langle [\boldmu^\ast-\boldmu_0]\, 
[\boldmu^\ast-\boldmu_0]^\dagger\rangle 
= \frac{1}{nN'}H^{-1} 
= \frac{1}{N}H^{-1} 
\ .
\label{A3}
\end{equation}

\section{Statistical uncertainties for Bell's
  Model}
\label{sigma_bell}

In order to determine the covariance matrix of the maximum likelihood
estimate for the basic model, in section \ref{s4a} the quantity
$s(\boldmu)$ as defined in (\ref{3.40}) had to be calculated.  Details
of this calculation are given below.

We first calculate
\begin{eqnarray}
  \label{A4}
  &&E(\alpha) := \langle e^{\alpha f} \rangle_1 = 
  \int_0^\infty df\; e^{\alpha f} p_1(f|\boldmu_0, v)\\
    \nonumber
  &&=
  \int_0^\infty df e^{\alpha f}
  \frac{e^{\lambda_0 + \alpha_0 f}}{r} 
  \exp\left(-\frac{e^{\lambda_0}}{r}
    \frac{e^{\alpha_0 f}-1}
    {\alpha_0}\right)
\end{eqnarray}
for an arbitrary loading rate $r=\kappa v$ and $\alpha>-\alpha_0$.  It
is convenient to rewrite equation (\ref{A4}) using the dimensionless
quantities $\tau := e^{\lambda_0}/(r\alpha_0)$ and $\eta =
\alpha/\alpha_0$ and to substitute $t=\tau \exp(\alpha_0 f)$ yielding:
\begin{equation}
  \label{A5}
  E(\alpha) = \tau^{-\eta} e^\tau 
  \int_\tau^\infty dt\; t^\eta e^{-t}
  \ .
\end{equation}
Using that $\tau \ll 1$ in typical AFM pulling experiments, we derive
at
\begin{equation}
  \label{A6}
  E(\alpha) = \tau^{-\eta} \Gamma(\eta+1) 
  + {\mathcal{O}}(\tau),
\end{equation}
where $\Gamma(\cdot)$ denotes the Gamma function.  Equation (\ref{A6})
directly gives the expected rupture force:
\begin{eqnarray}
  \label{A7}
  \langle f \rangle_1
  =
  \frac{\partial}{\partial \alpha} E(\alpha=0) 
  = -\frac{1}{\alpha_0}
  \left({\mathcal{C}} +\ln\tau \right) + {\mathcal{O}}(\tau)
\end{eqnarray}
with ${\mathcal{C}} \approx 0.577$ the Euler constant.  Using
equations (\ref{A6} \ref{A7}) and the linearity of the expectation
value we obtain:
\begin{eqnarray}
\nonumber
&&- \langle \ln p_1(f| \boldmu, r) \rangle_1 = \\
\label{A8}
&&-\lambda + \ln(\kappa v) 
+ \eta ({\mathcal{C}} + \ln \tau)
+\frac{e^\lambda}{\kappa v \alpha}\frac{1}{\tau^\eta}
\Gamma(\eta+1)
\ .
\end{eqnarray}
Finally, equation (\ref{A8}) together with the definition of the
quantity $s(\boldmu)$ yields the desired result (\ref{4.30}).

\section{Optimal bin-width for histograms}
\label{opt_bin}

Let $p_1$ be a probability density function with two continuous and
bounded derivatives.  For a sample of size $N$ the histogram estimate
$\hat{p}_1$ of $p_1$ is defined as
\begin{equation}
  \hat{p}_1(f) = \frac{\Lambda_N(f)}{Nh_N}
  \ ,
\label{A10}
\end{equation}
with $\Lambda_N(f)$ the number of values falling into the bin of width
$h_N$ around $f$.  Then one can show \cite{Sco79,Mat82} that for large
sample sizes the integrated mean squared error
\begin{equation}
  IMSE = \int df\; \langle 
  \left( \hat{p}_1(f) - p_1(f) \right)^2
  \rangle_1
\label{A11}
\end{equation} 
considered as a function of the bin-width is minimized by
\begin{equation}
  h_N^\ast = \left(
  \frac{6}
  {\int df\; (p_1'(f))^2}
  \right)^{1/3} N^{-1/3}
  \ .
\label{A12}
\end{equation}
Following the same lines as in Appendix \ref{sigma_bell} we obtain for
the Bell model:
\begin{equation}
  \int df (p_1'(f| \boldmu, v))^2 
  = \frac{1}{8} \alpha^3  
  \left( 1+ 
       {\mathcal{O}} 
       \left(\frac{2e^\lambda}{\kappa v \alpha }\right)  
  \right)
  \ .
  \label{A13} 
\end{equation}
Inserting (\ref{A13}) into (\ref{A12}) yields the optimal bin-width:
\begin{equation}
  h^\ast_N \approx 3.63 \frac{1}{\alpha N^{1/3}}
  \ .
  \label{A14}
\end{equation}
It should be noted that in the limit $\frac{2e^\lambda}{\kappa v
\alpha } \ll 1$, the optimal bin-width depends solely on the sample
size $N$ and on the parameter $\alpha$ which determines the width of
the distribution, but neither on the force-free dissociation rate $k_0
= \exp(\lambda)$ nor on the pulling velocity $v$.

If one wishes to determine the optimal bin-width prior to parameter
estimation, one may make use of the well known result $\langle (f -
\langle f \rangle_1 )^2\rangle_1 \approx \pi^2/(6\alpha^2)$ and choose
\begin{equation}
  h_N = 2.83 s_N N^{-1/3} 
  \ ,
  \label{A15}
\end{equation}
where $s_N$ denotes the standard deviation of the measured rupture
forces.  This estimate is pretty close to the optimal bin-width
$h_{N,Gauss} \approx 3.49 s_N N^{-1/3}$ of a Gaussian distribution
\cite{Sco79}.
\end{appendix}


\end{document}